\documentstyle[12pt,aaspp4]{article}
\makeatletter
\def\gnyoro{\mathrel{\mathpalette\gl@align>}}
\def\gl@align#1#2{\lower.6ex\vbox{\baselineskip\z@skip\lineskip\z@\ialign{$\m@th
#1\hfil##\hfil$\crcr#2\crcr\sim\crcr}}}
\makeatother

\begin{document}

\title{\bf PHYSICAL INTERPRETATION OF THE MASS-LUMINOSITY RELATION
OF DWARF SPHEROIDAL GALAXIES} 
\author{\bf HIROYUKI HIRASHITA$^{1,2}$, 
TSUTOMU T. TAKEUCHI$^{1,2}$, \\ AND NAOYUKI TAMURA$^1$} 
\affil 
{$^1$:  Department of Astronomy, Faculty of Science, Kyoto University,
Sakyo-ku, Kyoto 606-8502, Japan}
\affil
{$^2$:  Research Fellow of the Japan Society for the Promotion of
Science}
\centerline{Accepted by {\it ApJL}}
\centerline{E-mail: hirasita@kusastro.kyoto-u.ac.jp}
\authoremail{hirasita@kusastro.kyoto-u.ac.jp}
\begin{abstract}

We discuss a physical interpretation of the relation between $M_{\rm
vir}/L$ and $M_{\rm vir}$ of dwarf spheroidal galaxies (dSphs), where
$M_{\rm vir}$ and $L$ are the virial mass and the total luminosity of
a dSph, respectively. We used 11 dSphs in the Local Group as the
sample. We find two distinct sequences on the
$M_{\rm vir}/L$--$M_{\rm vir}$ plane:
$M_{\rm vir}/L\propto M_{\rm vir}^{2.0}$ for dSphs with
$M_{\rm vir}<10^8M_\odot$, whereas $M_{\rm vir}/L\simeq$ constant for
$M_{\rm vir}>10^8M_\odot$. A
``discontinuity'' is seen at $M_{\rm vir}\simeq 10^8M_\odot$.
We interpret the ``discontinuity'' as the threshold for the gas in
dSphs to be blown away by successive supernovae. If a dSph has
virial mass (most of which is dark mass) less than $10^8M_\odot$,
the gas is blown away, while in a dSph of larger mass, the deep
potential well prevents the blow-away mechanism from working
effectively. Thus, large mass ratio of dark
matter (DM) to baryonic matter (i.e., large $M_{\rm vir}/L$) is
realized in a
low-mass ($M_{\rm vir}<10^8M_\odot$) dSph through the gas depletion,
whereas $M_{\rm vir}/L$ becomes lower in high-mass
($M_{\rm vir}>10^8M_\odot$) dSphs. We further make an attempt to
explain the above relation for the
low-mass dSphs, $M_{\rm vir}/L\propto M_{\rm vir}^{2.0}$, based on
estimate of cooling time, using the scaling laws of virial
temperature, virial mass and radius of a dSph and
assuming that the heating by OB-star radiation terminates the star
formation activity. We succeed in deriving the above relation for
the mass-to-light ratio and luminosity.

\end{abstract}


\section{INTRODUCTION}

The Galaxy and M31 are surrounded by small companion galaxies.
Some of them belong to the classification
category called ``dwarf spheroidal galaxies (dSphs).''
Through recent observations, we have been obtaining the properties of
the Local Group dSphs. The dSphs have luminosities
of order $10^5$--$10^7\, L_\odot$, and are characterized by their
low surface brightnesses (Gallagher \& Wyse 1994 for review).
The knowledges of such low-luminosity objects are
important for some reasons. For example, in a galaxy formation
theory based on the cold dark matter (CDM) model, low mass galaxies
are considered to be the first bound luminous objects
(e.g., Blumenthal et al. 1984). Thus, the dSphs are expected to have
dark matter (DM).

Observations of velocity dispersions of dSphs (Aaronson 1983; see
also Mateo et al. 1993 and references therein) indicate too large
virial mass to be accounted for by the visible stars in the dSphs.
In other words,
dSphs have generally high mass-to-luminosity ratio. This fact can
imply the presence of
DM in these systems (e.g., Mateo et al. 1993).
The large spatial distribution of stars to their outer
regions (Faber \& Lin 1983) also suggests the existence of DM.

However, if a dwarf galaxy orbiting the Galaxy is
significantly perturbed
by Galactic tides, observed velocity dispersion of the dwarf
can be larger than gravitationally equilibrium dispersion (Kuhn \&
Miller 1989; Kroupa 1997).
Moreover, Bellazzini, Fusi Pecci, \& Ferraro (1996) pointed out
that there is a strong correlation between tidal forces by the
Galaxy and the surface brightnesses of the dSphs and that this
correlation supports the strong tidal perturbation by the Galaxy.
Hirashita, Kamaya, \& Takeuchi (1998, Paper I), however, showed
that this correlation is not exclusively supports the tidal picture,
since it may reflect the correlation between the galactocentric
distances and the surface brightnesses of the dSphs.
Paper I seem to support the DM-dominated picture
of the dSphs. Piatek \& Pryor (1995) and Oh, Lin, \& Aarseth (1995)
showed by numerical simulations
that it is difficult to inflate the central velocity dispersion of
dSphs by tidal force and to account for the apparent large virial mass
calculated from the observed velocity dispersions.

The dSphs are known to contain such a small amount of gas that
they show little evidence of recent star formation (e.g., Gallagher
\& Wyse 1994). Saito (1979a) showed that instantaneous gas ejection 
from supernovae (SNe) can make the gas in proto-dSphs
escape (see also Larson 1974).
However, he considered only self-gravity of gas: the DM potential,
whose existence is both observationally and theoretically supported
as mentioned above,
is not taken into account. In the presence of DM potential well, the
gas may not easily escape (Mac Low \& Ferrara 1998, hereafter MF98;
Ferrara \& Tolstoy 1998, hereafter FT98).
Dekel \& Silk (1986) showed that the SN feedback mechanism nicely
accounts for the observed scaling relations of mass, luminosity and
metallicity of each dSph, taking into account the presence of DM
halo (see also Peterson \& Caldwell 1993).

Since recent observations of the Local Group dSphs provide us
data with much better quality than those in ten years ago, we
re-examined the relation of observed quantities of the dSphs.
We will consider initial star formation (SF) and SN
feedback in the formation epochs of the dSphs with the DM-dominated
picture, in order to derive the present
relation between their virial mass and luminosity. Total
bolometric luminosity and virial mass of a dSph are denoted as $L$
and $M_{\rm vir}$, respectively. First of all, in the next section,
we present the correlation
between $M_{\rm vir}/L$ and $M_{\rm vir}$ for the Local Group
dSphs. The condition for the blow-away by SNe is explained in \S 3.
We present a physical interpretation of the
$M_{\rm vir}/L$--$M_{\rm vir}$ relation in
\S 4. The final section is devoted to the summary.

\section{$M_{\rm vir}/L$--$L$ RELATION}

In this section, we present the observed properties of the Local
Group dSphs. We focus on their virial mass
$M_{\rm vir}$ and total bolometric luminosity $L$. The virial mass
is derived from the internal velocity dispersion by using the
King-model fitting (Irwin \& Hatzidimitriou 1995). We used the
compiled data in
Table 11 of Mateo et al. (1993) for the virial masses and the total
luminosities for Draco, Carina, Ursa Minor, Sextans, Sculptor,
Fornax, NGC 147, NGC 185, and NGC 205. For Leo I, we used the
data in Zaritsky et al. (1989), Caldwell et al. (1992), and
Lee et al. (1993), and for Leo II, we refer to
Vogt et al. (1995) (see also Table 1 of Paper I).

Figure 1 shows the relation between $M_{\rm vir}/L$ and
$M_{\rm vir}$ for the above 11 dSphs. We divide these galaxies into
two groups; the lower-mass group (Group L; Draco, Carina,
Ursa Minor, Sextans, Sculptor, Leo I and Leo II) and the
higher-mass group (Group H; Fornax, NGC 147, NGC 185, and
NGC 205). The virial mass of
the former group is less than $10^8M_\odot$, while that of the latter
is more than $10^8M_\odot$.

We see that $M_{\rm vir}/L$ of Group H is almost
constant, though the number of samples is small. Analyses with larger
number including samples out of the Local Group show that bright
dwarf spheroidals (dwarf ellipticals)\footnote{In this {\it Letter},
we do not distinguish between dwarf spheroidals and dwarf
ellipticals.} are on a distinct sequence from Group L
(Bender, Burstein, \& Faber 1992; Peterson \& Caldwell 1993)
though the
dispersion of $M_{\rm vir}/L$ is large (an order of magnitude).

As for Group H, the constant $M_{\rm vir}/L$ may be due to the
inefficiency of escape of SN-heated gas (Dekel \& Silk 1986;
MF98; FT98), since the
potential of DM is deep.
This is discussed in the next section. On the other hand, for
galaxies belonging to Group L, the gas easily escapes out of
them once their gas is heated by SNe
(Saito 1979a) and OB-star radiation, because of
their shallow gravitational potentials. We discuss this point
later in \S 4. We note that the two groups can be distinguished
in the $\log M_{\rm vir}/L$--$\log L$ plot (Figure 8 of
Vogt et al. 1995). Our plot in Figure 1
($\log M_{\rm vir}/L$--$\log M_{\rm vir}$ plot) makes the
separation of the groups prominent.

The log-linear regression for the quantities in Group L is
\begin{eqnarray}
\log (M_{\rm vir}/L)=2.0\log M_{\rm vir}-13.1,\label{reg}
\end{eqnarray}
and the correlation coefficient is 0.92, which shows a strong
correlation. For the correlation study for other quantities
concerning the Local Group dSphs, see Bellazzini, Fusi Pecci, \&
Ferraro (1996), and Paper I. An interpretation of equation
(\ref{reg}) is discussed in \S 4.

\section{BLOW-AWAY CONDITION FOR PROTO-DWARFS}

As discussed by Saito (1979a), SN-driven winds blow the
gas in low-mass proto-galaxies away because of their
shallow potential wells.
Recent numerical simulations of SN-driven wind at initial
starburst of low-mass galaxies (MF98;
FT98) showed that if their
gas mass is more than $10^7M_\odot$, the gas ejection
efficiency is very low. This mass corresponds to $M_{\rm vir}
=10^8M_\odot$ if the mass fraction of baryon is 0.1.
This value agrees with the separation line of the two groups
(H and L; Figure 1).

This fact is physically interpreted as follows. Galaxies
with $M_{\rm vir}>10^8M_\odot$ form stars whose total mass is
proportional to the initial gas mass, which leads to the
constant mass-to-light ratio (Dekel \& Silk 1986). On the other
hand, galaxies with $M_{\rm vir}<10^8M_\odot$
blow the gas away soon after the formation of the
first-generation stars:
If a little fraction of gas becomes stars,
the SNe resulting from these stars are enough to blow away
the rest of the gas (Saito 1979a; Nath \& Chiba 1995). Thus, the
members in the low-mass category tend to have little baryonic
matter (i.e., higher $M_{\rm vir}/L$).

\section{LOW MASS DWARF SPHEROIDALS}

Equation (\ref{reg}) shows that the virial mass and total luminosity
of the lower-mass group (Group L) in Figure 1 satisfy the
relation, 
\begin{eqnarray}
\frac{M_{\rm vir}}{L}\propto M_{\rm vir}^{2.0\pm 0.79}.
\label{obs}
\end{eqnarray}
The error of the index is obtained from the uncertainty of
$\log M_{\rm vir}$ (typically $\sim 0.70$; estimated from the
uncertainty of galactocentric distance and velocity dispersion).
In this section, we will derive this relation by considering
the physical processes in the formation epoch.

The presence of DM in dSphs is indicated by observations of
velocity dispersions (e.g., Mateo et al. 1993; but see Kuhn \&
Miller 1989). Thus, it is reasonable to consider the gas
collapse in DM potential to form proto-dSphs. We assume
that the distribution of DM is not affected by baryon, since
the mass fraction of baryon is much lower than DM
(e.g., Bahcall 1997).

We assume that the initial star formation rate (SFR) of a dSph is
determined by the cooling time $t_{\rm cool}$ ($t_{\rm cool}\gnyoro
t_{\rm grav}$,
where $t_{\rm grav}$ is the free-fall time $\simeq 1/
\sqrt{G\rho_{\rm DM}}\sim 10^7$ yr).
We, here, note that the free-fall time is determined by the DM
mass density $\rho_{\rm DM}$, while the cooling time is determined
by the gas density.
The cooling time $t_{\rm cool}$ is
determined by the density and temperature of the gas as
\begin{eqnarray}
t_{\rm cool}\propto\frac{T}{\rho_{\rm gas}\Lambda (T)},\label{cool}
\end{eqnarray}
where $\Lambda (T)$ is the cooling function. Since the virial
temperature of such a low mass galaxy is much lower than $10^6$ K,
the cooling is dominated by H and He recombination (Rees \& Ostriker
1977): $\Lambda (T)\stackrel{\propto}{\sim}T^{-1/2}$
(Peacock \& Heavens 1990). From the above assumption that the SFR is
determined by the cooling
timescale, we obtain the following expression for the SFR
($\dot{M}$):
\begin{eqnarray}
\dot{M}\propto t_{\rm cool}^{-1}\propto
\frac{\rho_{\rm gas}}{T^{3/2}}.\label{sfr}
\end{eqnarray}
The temperature in the quasistatic collapse phase is
determined by virial temperature (Rees \& Ostriker 1977).
 Thus, the following expression for the
temperature is satisfied:
\begin{eqnarray}
T\propto\frac{M_{\rm vir}}{R},\label{virial}
\end{eqnarray}
where $R$ is the
typical size of the DM distribution.
Using the scaling relation of virial mass and size (Saito 1979b;
Nath \& Chiba 1995);
$R\propto M_{\rm vir}^{0.55}$,
we obtain from relations (\ref{sfr}) and (\ref{virial})
\begin{eqnarray}
\dot{M}\propto M_{\rm vir}^{-1.33},\label{scale}
\end{eqnarray}
where we assume that the initial mass ratio of gas to DM is
constant (i.e., $\rho_{\rm gas}\propto M_{\rm vir}/R^3\propto
M_{\rm vir}^{-0.65}$.)

In normal galaxies, the SF is stopped in the epoch of onset of
galactic wind, when the thermal energy produced by SNe
becomes equal to the binding energy of the galaxy (Arimoto
\& Yoshii 1987). However, in dwarf galaxies belonging to the
low-mass group, the heating by UV photons from first-generation OB
stars is large enough to supply the thermal energy equal to the
gravitational potential ($\sim 10$ OB stars $\sim 1$ OB
association). Thus the remaining gas in a low-mass dwarf
evaporates from the system soon after the formation of
first-generation stars.
We estimate the SF-terminating time by using
the crossing time of the wind generated
in an OB association. The velocity of the wind is estimated by the
sound speed $c_{\rm s}$ of $10^6$ K, which is a typical temperature
of the heated gas. The crossing time $t_{\rm cross}$ is estimated
as $t_{\rm cross}\simeq R/c_{\rm s}$. If we keep $c_{\rm s}$
constant, the following expression for the
$t_{\rm cross}$ is obtained by using the above scaling law
$R\propto M_{\rm vir}^{0.55}$:
\begin{eqnarray}
t_{\rm cross}\propto M_{\rm vir}^{0.55}.\label{cross}
\end{eqnarray}
We note that the typical value for $t_{\rm cross}$ becomes
$10^7$ yr (for $R=1$ kpc and $c_{\rm s}=100$ km s$^{-1}$), which is
shorter than the cooling time.
Thus, the formation of the second-generation stars is difficult.
The total mass of the stars, $M_*$, is estimated by
$\dot{M}t_{\rm cross}$. Thus, from relations (\ref{scale}) an
(\ref{cross}), we obtain
\begin{eqnarray}
M_*\propto M_{\rm vir}^{-0.78}.
\end{eqnarray}
This relation means that the mass-to-light ratio is scaled as
\begin{eqnarray}
\frac{M_{\rm vir}}{L}\propto M_{\rm vir}^{1.78},\label{theor}
\end{eqnarray}
where $L$ is considered to be proportional to $M_*$.

Comparing the observed relation (\ref{obs}) with the theoretical
prediction (\ref{theor}),
we see that the two agree well in the range of the observational
error. We note that it is difficult to obtain an absolute
value for the mass of the formed stars, since the star formation
efficiency is uncertain. Thus, we here only calculate the
scaling relation.

\section{SUMMARY}

We discuss a physical interpretation of the relation between $M_{\rm
vir}/L$ and $M_{\rm vir}$ of dwarf spheroidal galaxies (dSphs), where
$M_{\rm vir}$ and $L$ are the virial mass and the total luminosity of
a dSph, respectively. We used 11 dSphs in the Local Group as the
sample. We find two distinct sequences on the
$M_{\rm vir}/L$--$M_{\rm vir}$ plane:
$M_{\rm vir}/L\propto M_{\rm vir}^{2.0}$ for dSphs with
$M_{\rm vir}<10^8M_\odot$, whereas $M_{\rm vir}/L\simeq$ constant for
$M_{\rm vir}>10^8M_\odot$. A
``discontinuity'' is seen at $M_{\rm vir}\simeq 10^8M_\odot$
(Figure 1). We
interpret the ``discontinuity'' as the threshold for the gas in
dSphs to be blown away by successive supernovae (MF98; FT98). If a
dSph has
virial mass (most of which is dark mass) less than $10^8M_\odot$,
the gas is blown away, while in a dSph with larger mass, the deep
potential well prevents the blow-away mechanism from working
effectively. Thus, large mass ratio of DM to gas (i.e., large
$M_{\rm vir}/L$) is realized in a
low-mass ($M_{\rm vir}<10^8M_\odot$) dSph through the gas depletion,
whereas $M_{\rm vir}/L$ becomes lower in high-mass
($M_{\rm vir}>10^8M_\odot$) dSphs.

We further explained the above relation for the
low-mass dSphs, $M_{\rm vir}/L\propto M_{\rm vir}^{2.0}$ based on
estimate of cooling time. We used scaling laws of virial temperature,
virial mass and radius of dwarf spheroidals.
The timescale for the duration of SF is estimated by the crossing
time of OB-association-heated wind, since the heating by
OB-star photons is effective in low-mass systems.
We finally note that this condition for the cessation of SF
in low-mass dSphs
is different from that in higher-mass galaxies where the
SN heating is essential to terminate the SF.

\acknowledgements
We would like to thank the anonymous referee for the useful comments
which improved our paper.
We thank Profs. M. Sait\={o}, S. Mineshige and
K. Ohta for their continuous encouragement. We are grateful to Drs.
A. Ferrara,
M.-M. Mac Low and E. Tolstoy for their kindly sending us
their preprints and helpful comments on
the starburst-driven mass loss. We also thank Dr. H. Kamaya for
fruitful discussions and invaluable comments. Two of us (HH and TTT)
acknowledge the Research Fellowship of the Japan Society for the
Promotion of Science for Young Scientists. We fully utilized the
NASA's Astrophysics Data System Abstract Service (ADS).


\newpage

\centerline{\bf FIGURE CAPTION}

\noindent
FIG. 1---
The relation between the mass-to-light ratio $M_{\rm vir}/L$ and
the virial mass $M_{\rm vir}$
of the Local Group dwarf spheroidal galaxies. The dotted line,
$M_{\rm vir}=10^8M_\odot$,
represents the dividing line of the two groups; high-mass group
(Group H) and low-mass group (Group L). The solid line is the
log-linear regression for the Group L.

\end{document}